\documentclass[sigconf, 10pt, nonacm=true, balance=false]{acmart}

\usepackage{listings}
\usepackage{color}
\usepackage{float}

\definecolor{codegreen}{rgb}{0,0.6,0}
\definecolor{codegray}{rgb}{0,0,0}
\definecolor{codepurple}{rgb}{0.58,0,0.82}
\definecolor{backcolour}{rgb}{0.95,0.95,0.92} 
\usepackage{xspace}

\newcommand{\sys}{\mbox{\textsc{ALTo}}\xspace}
 
\lstdefinestyle{mystyle}{
    backgroundcolor=\color{backcolour},   
    commentstyle=\color{codegreen},
    keywordstyle=\color{magenta},
    numberstyle=\tiny\color{codegray},
    stringstyle=\color{codepurple},
    basicstyle=\fontsize{8}{10}\ttfamily,
    breakatwhitespace=false,         
    breaklines=true,                 
    captionpos=b,                    
    keepspaces=true,                 
    numbers=left,                    
    numbersep=5pt,                  
    showspaces=false,                
    showstringspaces=false,
    showtabs=false,                  
    tabsize=2,
    frame=single
}
 
\def\BibTeX{{\rm B\kern-.05em{\sc i\kern-.025em b}\kern-.08emT\kern-.1667em\lower.7ex\hbox{E}\kern-.125emX}}

\begin{document}

%
\title{\sys: Ad Hoc High-Accuracy Touch Interaction Using Acoustic Localization }

%
\author{Arvind Seshan}
\affiliation{%
  \city{Fox Chapel}
  \state{PA}
}

%
\renewcommand{\shortauthors}{Seshan}

%

\begin{abstract}
 Millions of people around the world face motor impairments due to Parkinson’s, cerebral palsy, muscular dystrophy and other physical disabilities. The goal of this project is to increase the usable surface-area of devices for users with these disabilities by creating a simple, inexpensive, and portable way to enable high accuracy touch interaction with large surfaces such as a table or even a wall. 
 
 This project uses a novel approach that analyzes the acoustic signals at four piezoelectric microphones placed on the interactive surface to identify sounds related to the same event (e.g. a finger tap) at each of the microphones. \sys (\textbf{A}coustic \textbf{L}ocalized \textbf{To}uch) uses the results of this signal processing to compute the time difference of arrival (TDOA) across the microphones.  The collected TDOA data is used to compute an approximate location of a sound source (e.g., a finger tap) using a collection of hyperbolic equations. 
 
 An experimental evaluation of a system prototype was used to identify a number of software and signal processing optimizations needed to significantly improve accuracy and create a usable system. The results of the research indicate that it is possible to detect the location of a touch with high accuracy. The \sys prototype achieves an accuracy of 1.45cm in the x-direction and 2.72cm the y-direction which is within the range for the target usage (ie. those with motor impairments).
\end{abstract}

%
\keywords{time difference of arrival, TDOA, HCI, touch interaction, acoustic localization, piezoelectic discs, pseudo range multilateration, accessibility, motor impairment}

%
\maketitle

\section{Introduction}
\label{sec:intro}

Mobile devices are becoming more and more important to daily life. Recent studies estimate that
almost 80 percent of all Americans (including children) own a smartphone~\cite{smartphones}. These devices are also becoming more diverse; most recently smart watches have become increasingly popular (1 in 6 adults in the US own a smart watch)~\cite{smartwatch}. The design of these devices face two requirements that prove challenging to meet simultaneously. First, they must be small enough to be unobtrusive -- e.g., fit in a pocket or on a wrist. Second, touch remains the predominant user interface and these devices must be easy to interact with accurately. The result of these conflicting requirements, is that devices are often as small as possible while providing a usable touch surface. 

While the end result has been numerous popular and useful devices, the disappointing consequence of this design tradeoff is that they have become inaccessible for those with motor impairments due to illness or age. These users find it challenging to interact with such small surfaces~\cite{2014.Naftali} ~\cite{2017.Findlater}. Given the importance of these devices to professional (e.g. email), social (e.g. Facebook), and emergency (e.g. AMBER Alerts) communication, it is vital that these devices provide better accessibility to all users. 

Many mobile devices do provide a range of accessibility features, including screen magnification support, larger text/UI element sizing, and UI elements that are easier for users with motor impairment to use. However, none of these address the fundamental issue that existing device screens are too small for motor-impaired users to interact with. An alternative is to move to non-touch interfaces such as voice. While voice interaction has improved dramatically since the introduction of tools such as Apple Siri in 2011~\cite{siri}, voice controls are not appropriate for many tasks where touch dominates and voice controls cannot be used in many settings (e.g. in a classroom).

\begin{figure}[b]
  \centering
  \includegraphics[width=\linewidth]{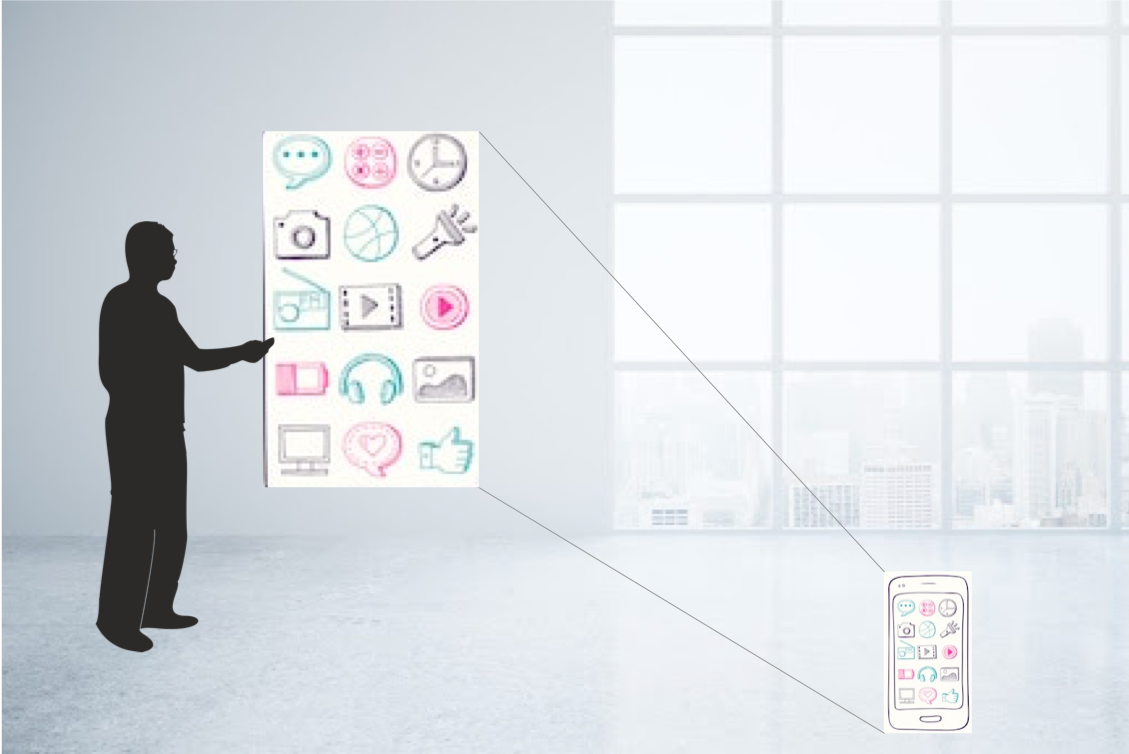}
  \caption{High-Level Vision for ALTo}
  \label{fig:ALToVision}
\end{figure}

The goal of this work is to address the fundamental issue of interactive surface size by making it possible to use any nearby surface (e.g., a desk, a wall or a chalkboard) for interaction. See Figure~\ref{fig:ALToVision}. This project introduces an acoustic approach to solving the problem. It applies acoustic time difference of arrival in a novel way to approximate the location of a user's tap on a large surface. This method allows for an inexpensive and portable solution to achieve high-accuracy touch localization on any surface. 

The study focuses on tapping because tapping is by far the most common touchscreen interactive technique \cite{2017.Findlater}. When a user taps on a surface, microphones pick up the sound. The time it takes for the sound to reach each microphone is directly related to the distance away from the sound source. This principal is essential to finding the location of the tap. The project analyses the difference in time between microphones to approximate the location of a tap by computing a collection of hyperbolic equations and solving for the intersection. 

An experimental evaluation of the ALTo (\textbf{A}coustic \textbf{L}ocalized \textbf{To}uch) system was used to identify a number of software and signal processing optimizations needed to significantly improve accuracy and create a usable system. The results of the research indicate that it is possible to detect the location of a touch with high accuracy.

The rest of this paper is organized as follows. Section~\ref{sec:system} describes the design for the ALTo system, including the system requirements and the methods used. Section~\ref{sec:software} focuses on the software design for both the data collection and analysis portions of the project. Section~\ref{sec:results} provides a detailed description of the results and an analysis of the data from this study.


\section{System Overview}
\label{sec:system}

In this section, I describe the design of \sys. I first describe the key requirements that any system designed to provide 
ad hoc touch interaction for mobile devices must address. Second, I discuss some of the underlying techniques that \sys relies on. 
Finally, I describe the details of the \sys design and implementation.

\subsection{System Requirements}
\label{sec:require}

\begin{figure}[t]
\centering
\includegraphics[width=\linewidth]{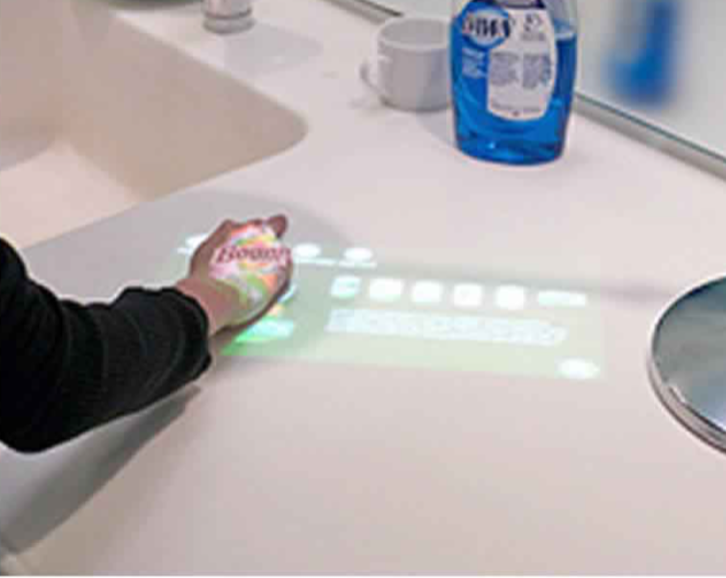}
\caption{Interactive Surface Sample. Image Source: ~\cite{DeskProjection} }
\label{fig:deskprojection}
\end{figure}

The goal of this system is to enable applications to use arbitrary surfaces as an input source much like a touch screen or mouse. This would allow for a wide range of application designs with larger interaction elements needed by motor-impaired users. For example, the application or device could project an image onto a surface for interaction (Figure~\ref{fig:deskprojection}). To enable such applications, whenever the user taps a surface, the \sys system must be able to determine the (x,y) coordinates of taps and pass this information on to the application. 

In addition to meeting this high-level goal, the design of the \sys system needs to meet the following important requirements:

\begin{itemize}
    \item {\bf Instrumentation.}  There should be no prior instrumentation of the surface required. Many existing approaches to enabling this type of interaction, such as Smart Rooms~\cite{smartroom} and gesture recognition systems~\cite{gesture}, require significant specialized infrastructure to be added to the room. This may include cameras, active sensors, or beacons. 
    
    \item {\bf Portable.} Any sensors or components needed by the system must be either integrated into the mobile device or be similar in size/portability as the mobile device. Given the battery life constraints of mobile devices, the solution should use little if any power. 
    
    \item {\bf Surface Support.} The user should be able to use a wide range of accessible surfaces, which may be made of a variety of materials.
    
    \item {\bf Accurate.} The system must have high accuracy. User taps on a surface should be localized with an error less than 2cm in any direction. This accuracy is based on the target use by motor impaired individuals. Previous studies indicate that those with motor impairments can have up to an 18mm error when using a touch screen \cite{2017.Findlater}. 
    \item {\bf Inexpensive.} The system components must not add too much expense to the device.
\end{itemize}

\subsection{Approach: Multilateration}

Based on these requirements, the project applies an approach known as multilateration. This technique has been used since World War II in navigation systems, but has been recently been replaced by the Global Positioning System (GPS). Multilateration is also used in air traffic management and surveillance systems~\cite{multilateration}. Muiltilateration uses the Time Difference of Arrival (TDOA), i.e. the different arrival times of a common source signal at each of three or more receivers, to locate the position of a signal source.

\begin{figure}[t]
  \centering
  \begin{tabular}{c|c}
      \includegraphics[width=.45\linewidth]{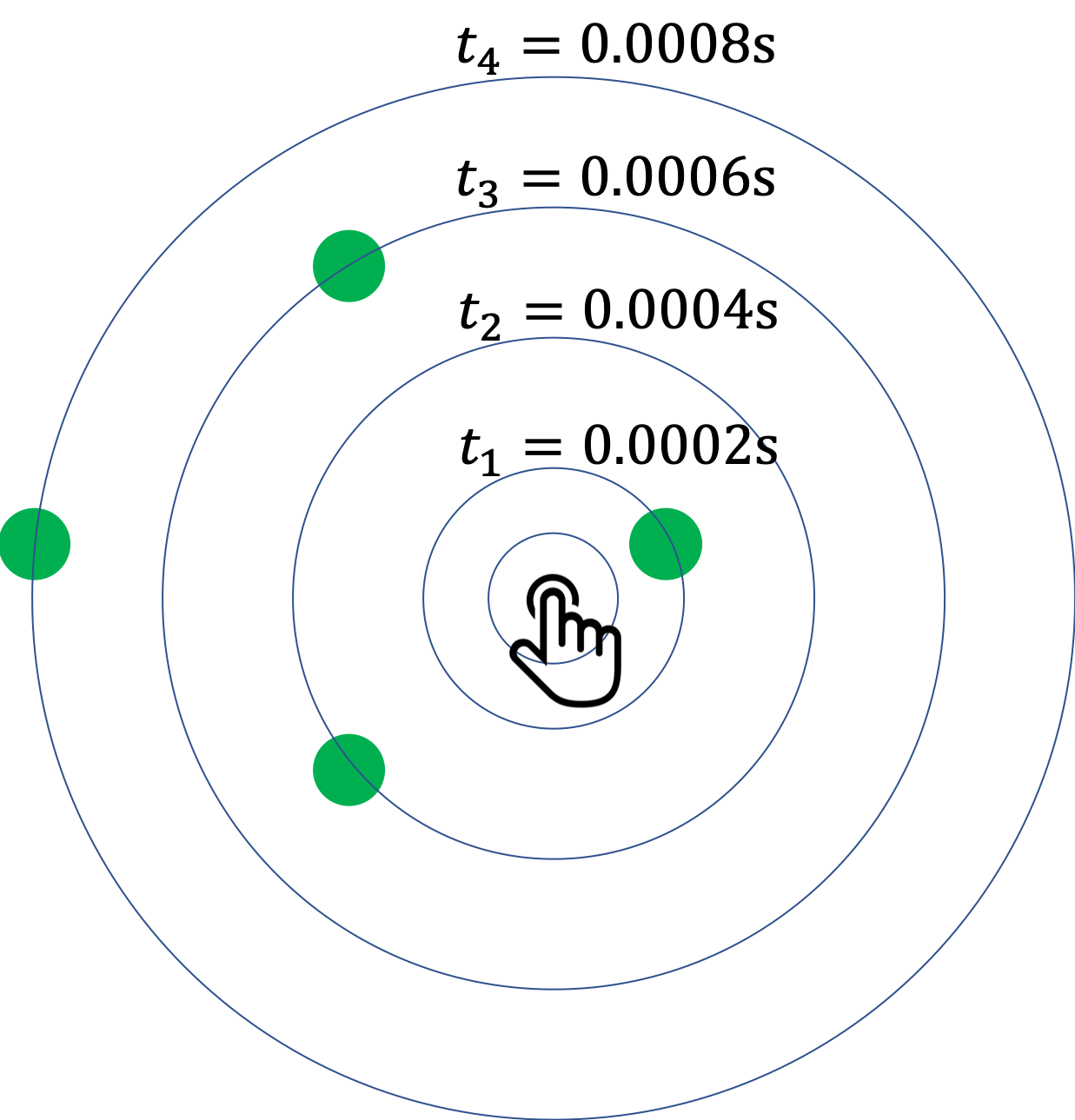}   &    \includegraphics[width=.45\linewidth]{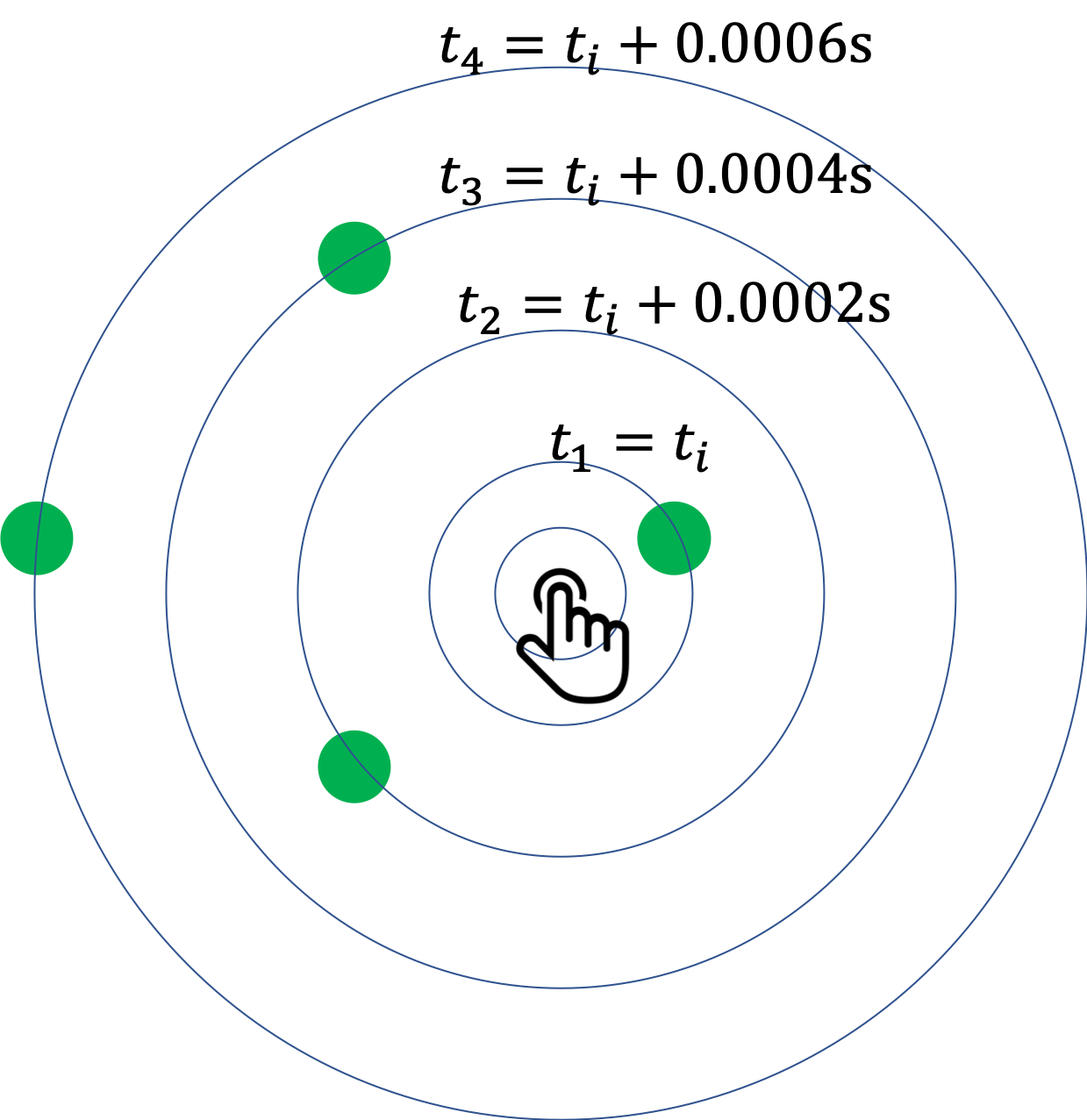} \\
       (a) & (b) \\
        \\
         \includegraphics[width=.45\linewidth]{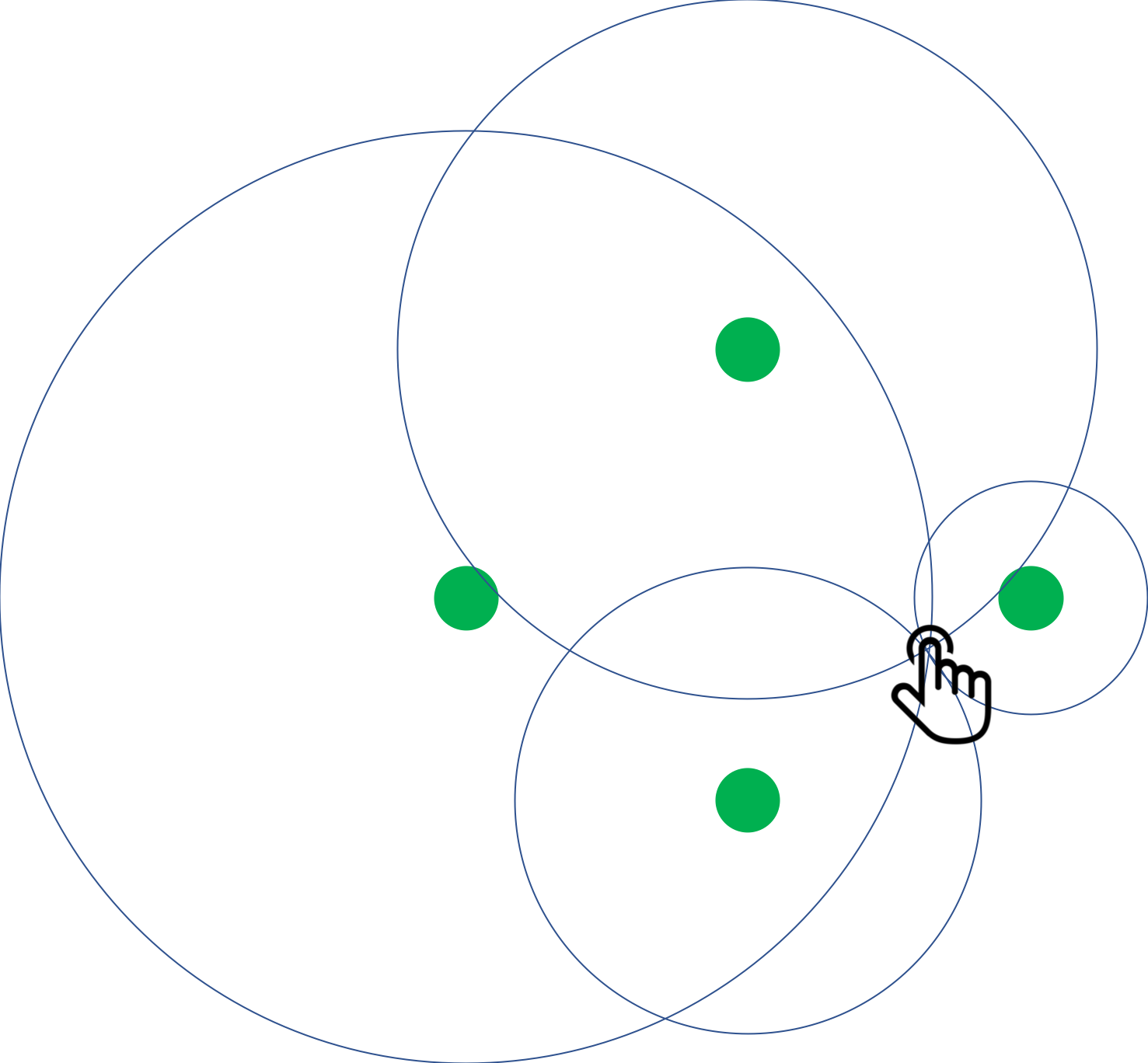} &  \includegraphics[width=.45\linewidth]{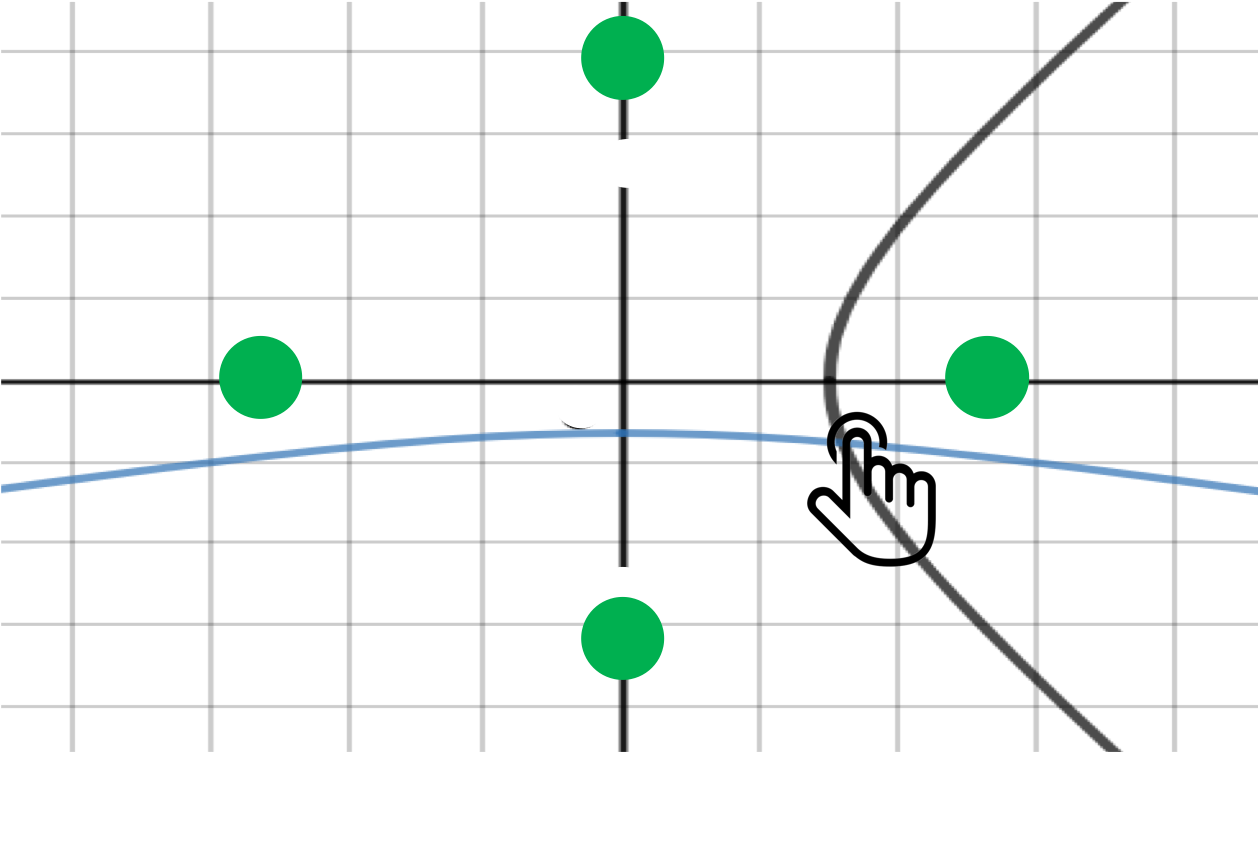} \\
       (c) & (d) \\
       \\
       \textbf{True Range} & \textbf{Pseudo Range} \\
       \textbf{Multilateration} &\textbf{Multilateration}
  \end{tabular}

  \caption{Illustration of multilateration techniques.}
  \label{fig:tdoa}
\end{figure}

Figure~\ref{fig:tdoa} illustrates the basic idea behind multilateration. If the user creates a signal at the point indicated by the finger, the signal will propagate outward from that point in all directions at a constant speed. In my system, the user creates the signal by tapping a surface and the signal itself is the acoustic wave that moves outward from this tap location through the material of the surface. The movement of the signal is represented by the concentric circles centered at the signal source. Each circle represents the locus of points at which the signal is located after a particular delay after the signal was generated. The delay is indicated as $t_1, t_2, t_3$ and $t_4$. On the part (a) of the figure, the labels show the absolute time after the tap was generated (i.e. time of the tap is 0 seconds). The green points represent locations where the signal is observed by some type of receiver, such as a sensor. If the absolute delay for the arrival of the signal at each sensor and the speed of propagation of the signals is known, it is easy to compute the distance from each sensor to the origin of signal. This implies that the origin must be located on the circles drawn around each sensor shown in part (c) of the figure. Finding the intersection of these circles identifies the precise location of the signal origin. This approach is known as true range multilateration. This approach requires knowing the precise distance from each of the sensor to the signal origin, which was determined based on the absolute time after the signal was generated. However, in practice, the system will not know when the user signal (in this system, when the user taps a surface) was generated. 

In \sys, the first observation of a signal is not when the tap is generated but when the signal reaches the closest sensor. In Figure~\ref{fig:tdoa}(b), the rightmost sensor receives the signal at some time $t_i$. The signal still propagates as it did in Figure~\ref{fig:tdoa}(a), reaching each of the sensors at the appropriate times. However, in this case each sensor simply observes when it receives the signal relative to the time that the rightmost sensor received the signal. The difference in time between any two sensors receiving the signal is equal to the difference in distance to the signal source divided by the speed of the signal. From this observation, any pair of sensors can determine that the signal origin must be located along the locus of points on the surface that have the computed distance difference between the sensors. The locus of points with a constant distance difference from to points in space is known as a hyperbola. Part (d) of the figure shows the hyperbolas that would be computed indicating possible locations relative to the top-bottom sensor pair and right-left sensor pair. Note only one half of a traditional mathematical hyperbola is shown since the system knows that the signal source is closer to one of the sensors in a pair due to the difference in delay that is measured. In addition, the figure has only two hyperbolas drawn; however, the hyperbolas between any sensor pair can be drawn based on the measurements shown in part (a) of the figure. There are, in fact, six hyperbolas that can be drawn using the 4 sets of sensors. In general, $n$ sensors would give information to generate $n(n-1)/2$ total hyperbolic equations to use. As with true range multilateration, the location of the signal source can be computed by determining the intersection of the possible locations (i.e. the intersection of the hyperbolas in this case). Note that while the drawings in Figure~\ref{fig:tdoa} show the tap location within the set of sensors, the approach can localize taps outside the set of sensors as well. This approach to localization is known as pseudo range multilateration (aka TDOA multilateration or hyperbolic navigation). \sys uses pseudo range multilateration. 

\subsection{\sys Hardware Design}
\label{sec:hardware}

The \sys hardware design must address a few key issues. The first of which is how to collect the sensor observations needed for pseudo range multilateration while addressing the requirements described in Section~\ref{sec:require}. 

The key components of the system are just the four piezoelectric disks to detect sound propagation through a hard surface. Piezoelectric disks are particularly good hardware for this project because they are inexpensive and compact. In addition, they are relatively easy to attach to any surface with some light adhesive and collecting observations from these sensors incurs little additional energy cost to the mobile device. 

\begin{figure}[t]
  \centering
  \includegraphics[width=\linewidth]{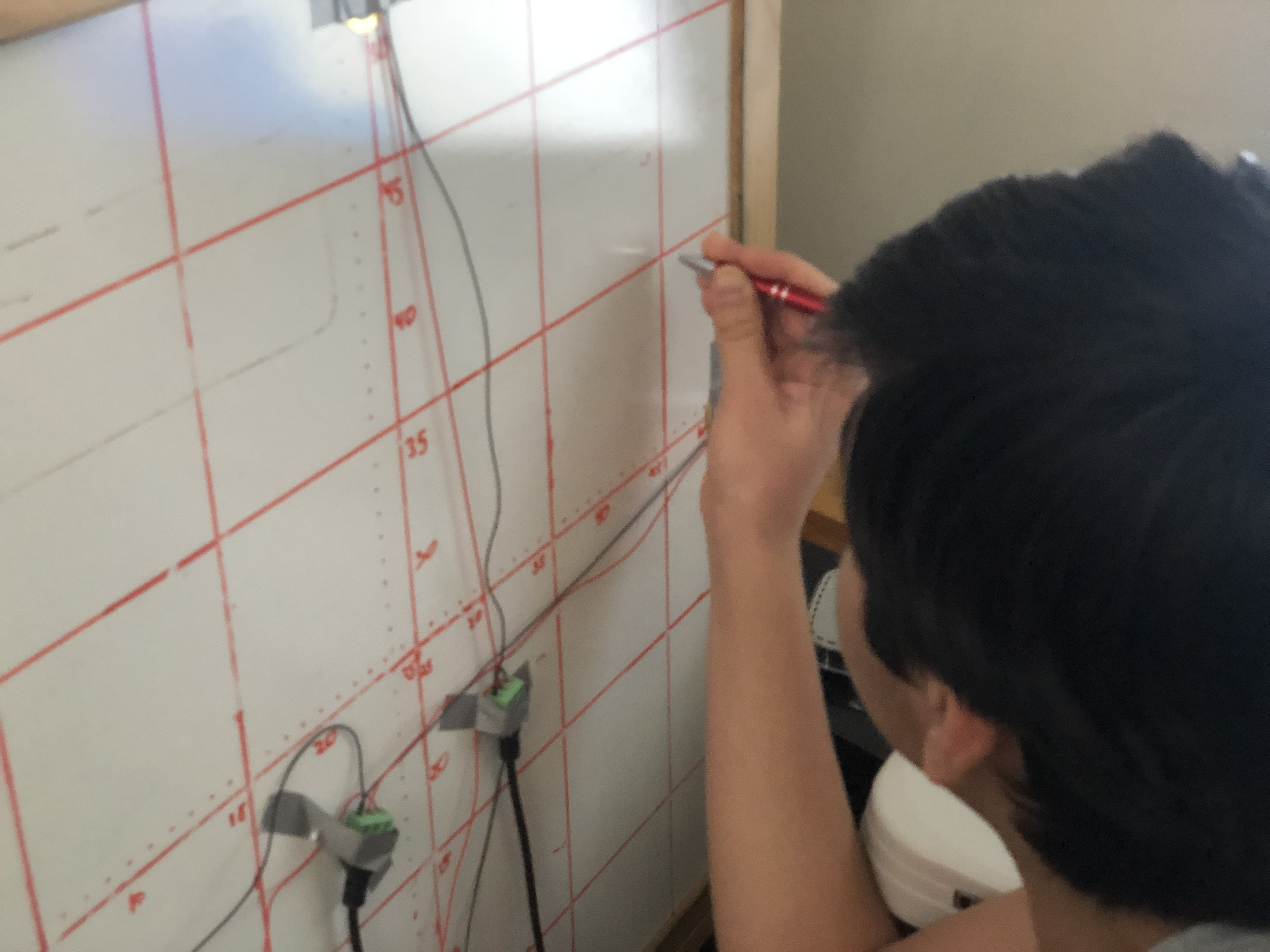}
  \caption{ALTo hardware used in the prototype}
  \label{fig:arvindtapping}
\end{figure}

The piezoelectric disks are connected to the microphone input of a computing device. Four disks require the use of two stereo inputs. While most mobile devices do not provide 2 sets of stereo inputs, the cost of such inputs is relatively small. In my experimental testbed, I use a standard desktop computer which has both a stereo microphone and line-in input for its standard motherboard audio interface. To connect the piezoelectric two wire output to the standard 3.5mm audio plug available on most computing devices, I cannibalized a pair of earphones and soldered their ear-plug connecting wires to the piezoelectric outputs. In later prototypes, I replaced the earphones with a 3 screw terminal to 3.5mm headphone jack converter to provide a more secure connection. (See Figure~\ref{fig:arvindtapping})

The above hardware design addresses several of the system requirements of being portable, inexpensive and easy to implement. However, the key requirement that is not clearly addressed is accuracy, especially across a wide range of surfaces. I address this concern through the project. The main challenge in addressing accuracy is that the time difference of arrival measurements made by the sensors must be precise. The speed of sound in air (at temperature 32C) is approximately 1260 km/h or 350 m/s. However, the speed of sound in materials such as wood is much faster, 3500 m/s, with some variation depending on the type of wood. This means that the sound signal can travel 1cm in a surface in as little as 1/350000 s or 3$\mu$s. The measurements from the system need to be accurate within this time range to provide the level of accuracy desired. This requires careful software design which I describe in Section~\ref{sec:software}.



    
 




\begin{figure}[t]
  \centering
  \includegraphics[width=\linewidth]{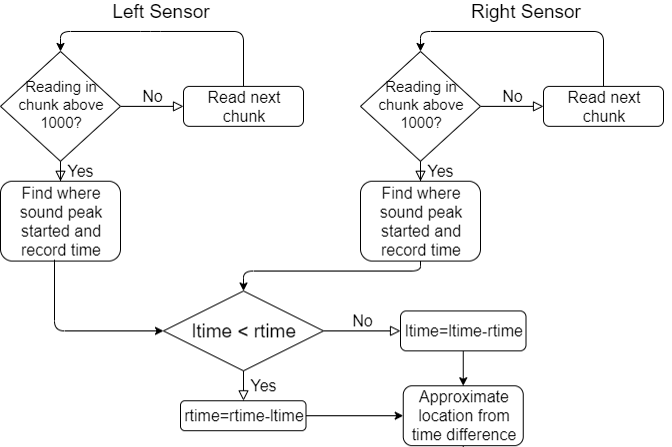}
  \caption{Flowchart representing the high level software structure for data collection with a single sensor.}
  \label{fig:singlesensor}
\end{figure}

\begin{figure*}[t]
  \centering
  \includegraphics[width=\textwidth]{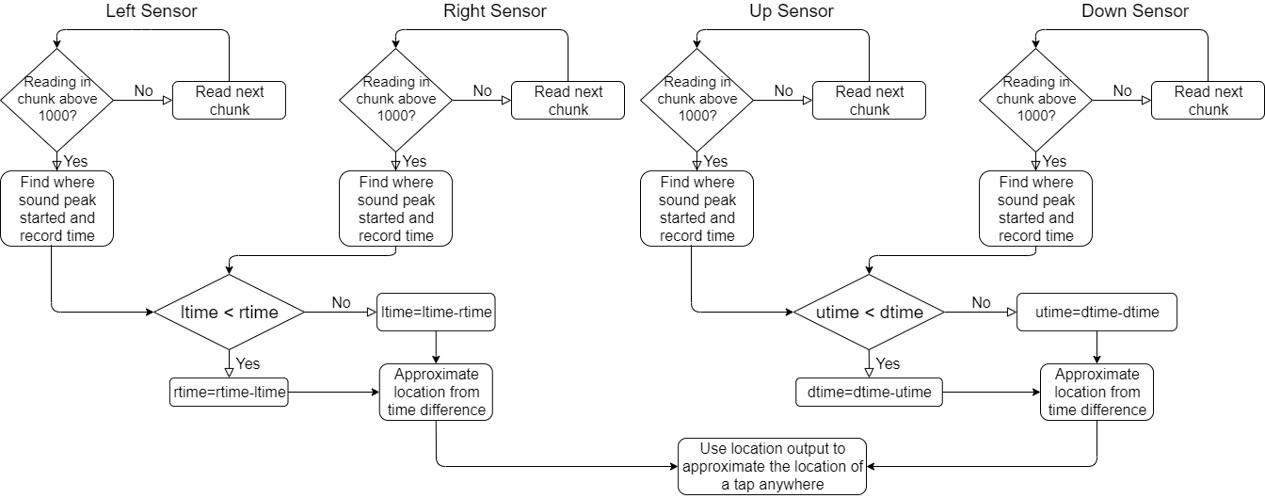}
  \caption{Flowchart representing the high level software structure for the two dimensional location estimate.}
  \label{fig:dualsensor}
\end{figure*}

\section{\sys Software Design}
\label{sec:software}

There are two major parts of the \sys software -- data collection and analysis. 

\subsection{Data Collection}

The Flowchart in Figure~\ref{fig:singlesensor} provides a high-level view of the code used to collect the needed sensor readings from a pair of piezoelectric microphones. The code (listed in Appendix~\ref{sec:code}) to implement this data collection is written in Python3 and uses the pyaudio library to collect and process audio data. Using pyaudio, the audio interface is configured with a sample rate and a buffer size. A read from the audio device completes when the buffer is full at which point the system returns a \textit{chunk} of audio sample data for the left and right audio channels (i.e. the two piezoelectric microphones). This is where the flowchart begins with the step "Read next chunk". The audio samples in the chunk are encoded as 16-bit values representing the amplitude of the audio signal at that time in the recording. A value above 1000 for the sample amplitude indicates a loud noise such as a tap on the surface. The code in the function hit\_test() iterates over the samples of a chunk looking for value above 1000. This is done for both the data from the left and right audio channels. If a loud noise was found in both audio channels, the sample number of the start of the loud sound is recorded for each channel and compared. Depending on which channel detected the tap noise earlier, the difference in sample number is computed (as either left\_sample\_number - right\_sample\_number or right\_sample\_number - left\_sample\_number). This difference in sample number is converted to a time difference by dividing by the configured sampling rate of the audio device. 

Extending this data collection to use four sensors adds some additional complexities. As shown in Figure~\ref{fig:dualsensor}, I begin by replicating the data collection workflow from the left and right sensor to the top and bottom sensor. Each pair of sensors is on a different audio device. As a result, the reading and processing of chunks is done independently. For example, the left and right sensors read of audio data always completes at the same time and their chunks are processed together. Similarly, the top and bottom sensor reads also complete simultaneously and are processed together. However, there is no relationship between the timing of left/right to top/bottom data read completion times. The implication of this is that the time difference computation between the matching pair of sensors (e.g. left and right) based on a natural common synchronized timing -- the sample number from the associated audio interface. However, to compare the time difference from a signal between a non-matching pair (e.g. right and up sensors) requires that the system compensates for the timing offset of their different audio device reads. I found that the accuracy requirements of the system made this impractical. As mentioned in Section~\ref{sec:hardware}, \sys requires accuracy on the order of $\mu$s. Introducing cross-device synchronization introduces too much error to meet this accuracy goal. As a result, I chose to rely solely on the time differences computed by paired sensors as part of tap localization.

Another challenge associated with the four sensor configuration was supporting concurrent processing of the independent sensor feeds. There were four different designs tested for this purpose:
\begin{itemize}
    \item \textbf{Sequential.} I began by using sequential reads of the audio device, relying on the buffering associated with the pyaudio interface to accommodate sequential processing of the data feeds. Unfortunately, this proved too slow and audio samples were dropped by the system, resulting in a non-working system. 
    
    \item \textbf{Python Threads.} The data collection system was tested using a single python thread for each audio device. This would allow the devices to collect and process data concurrently and still share data in a light-weight fashion. Unfortunately, Python's Global Interpreter Lock (GIL)~\cite{GIL} prevents Python threads from making use of multiple CPU cores. As a result, the delays between getting a particular thread scheduled for execution were significant. This also resulted in a non-working system design. It may have been possible to use more recent Python extensions for multiprocessing support; however, I choose to consider other options.
    
    \item \textbf{Asynchronous Callbacks.} The pyaudio system supports asynchronous callbacks to perform processing when audio data was ready. This allows light-weight multiplexing of audio processing. However, while this performed better than the sequential implementation, it also suffered from performance problems that made it non-working. 
    
    \item \textbf{Multiple Processes.} The fourth approach was to use multiple independent Python processes -- one per audio device used. Since these are independent processes they can run on concurrently on independent CPU cores - eliminating any issues with scheduling or resource contention. The key challenge was sharing the output from the two devices to compute an output coordinate. This could be done using a variety of inter-process communication methods, including sockets, shared memory, pipes or files. The goal of the current prototype is to show the feasibility of the approach, and, as a result, I choose to record the data and use an offline process to compute the tap coordinates.
    
\end{itemize}

\begin{figure}[t]
  \centering
  \includegraphics[width=\linewidth]{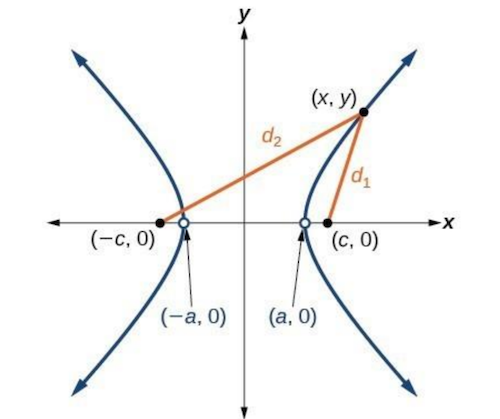}
  \caption{Sensor pair data analysis. Image Source: \cite{hyperbolas} }
  \label{fig:hyperbola}
\end{figure}

\subsection{Data Analysis}

Once a time difference between a pair of sensors is computed, \sys converts this to an distance estimate, $\Delta$, based on measurements of the speed of sound in the surface being used (Section~\ref{sec:speedsound}). Note that knowing the speed of sound is not necessary if applications do not need coordinate values in terms of normal units. For example, computations and calibrations can be made in terms of the size of a used region rather than absolute measurements. Here, I keep the discussion in terms of real units to simplify any descriptions. 

The distance $\Delta$ that is computed above is the difference in distance of the tap location relative to the two sensors. Figure~\ref{fig:hyperbola} helps illustrate this visually. The coordinate $(-c, 0)$ and $(c, 0)$ represent the coordinates of the two sensors. If the user tapped at location $(x, y)$, the distance to the two sensors would be $d_1$ and $d_2$. The distance $\Delta$ represents the value of $(d_2 - d_1)$. Note that there are many locations in the two-dimensional surface that would have this same difference in distance. The collection of such points describes the hyperbola shown in the Figure. For this particular hyperbola, $a = \Delta/2$ since the intercept with the x-axis is at $(a,0)$. 

In my prototype of \sys (described in Section~\ref{sec:speedsound}, the left and right sensors are located 26cm away from the origin. Given a measurement $\Delta$ and its corresponding x-axis intercept $a$, the equation describing the relevant hyperbola would be:

\begin{equation} \label{eq:1}
\frac{y^2}{(26-a)^2} - \frac{x^2}{(26^2 - (26 - a)^2)} = 1
\end{equation}
This simplifies to:
\begin{equation} \label{eq:2}
\frac{y^2}{(26-a)^2} - \frac{x^2}{(52a - a^2)} = 1
\end{equation}

The prototype also has top and bottom sensor located vertically 26cm away from the origin. Given a measurement $\Delta$ for the signal arrival between these sensors, \sys computes a corresponding y-axis intercept $b = \Delta/2$. This hyperbola is described by the equation:

\begin{equation} \label{eq:3}
\frac{x^2}{(26-b)^2} - \frac{y^2}{(26^2 - (26 - b)^2)} = 1
\end{equation}
This simplifies to:
\begin{equation} \label{eq:4}
\frac{x^2}{(26-b)^2} - \frac{y^2}{(52b - b^2)} = 1
\end{equation}

Solving these equations simultaneously, results in:

\scriptsize
\begin{equation} \label{eq:5}
x = \pm\frac{(a-26)\sqrt{b}\sqrt{-(-a^2b + 52a^2 + 52ab - 2704a + b^3 - 104b^2 + 3380b - 35152)}}{\sqrt{-(676a^2 - 35152a + 676b^2 - 35152b + 456976)}}
\end{equation}
\normalsize

\begin{equation} \label{eq:6}
y = \pm\sqrt{\frac{(-a^4 + a^2x^2 + 104a^3 - 52ax^2 - 3380a^2 + 35152a)}{(-a^2 + 52a - 676)}}
\end{equation}


Note that using the two hyperbola equations produces four intersections. To determine which intersection is the correct one, I looked at which sensors detected the tap first. The tap must be closer to the ones that detected the tap first. (e.g. if the top sensor and right sensor detected the tap before the bottom sensor and left sensor respectively, the tap must be in the intersection that is in Quadrant I). This produces a single $(x,y)$ that \sys can provide to programs to use. 





\section{Experimental Results}
\label{sec:results}

The goal of this experimental evaluation is to show that \sys can provide accurate $(x,y)$ coordinate information for applications to use. This done through a sequence of experiments that:
\begin{itemize}
    \item Show that a \sys can accurately measure the change in signal delays as a user's taps move from one sensor to another (Section~\ref{sec:1dtest})
    \item Show that adjustments to the sampling rate can improve accuracy (Section~\ref{sec:sampling})
    \item Compute the speed of sound in using a prototype (Section~\ref{sec:speedsound})
    \item Measure the $(x,y)$ accuracy of \sys (Section~\ref{sec:accuracy})
\end{itemize}

\begin{figure}[t]
  \centering
    \begin{tabular}{c}
  \includegraphics[width=\linewidth]{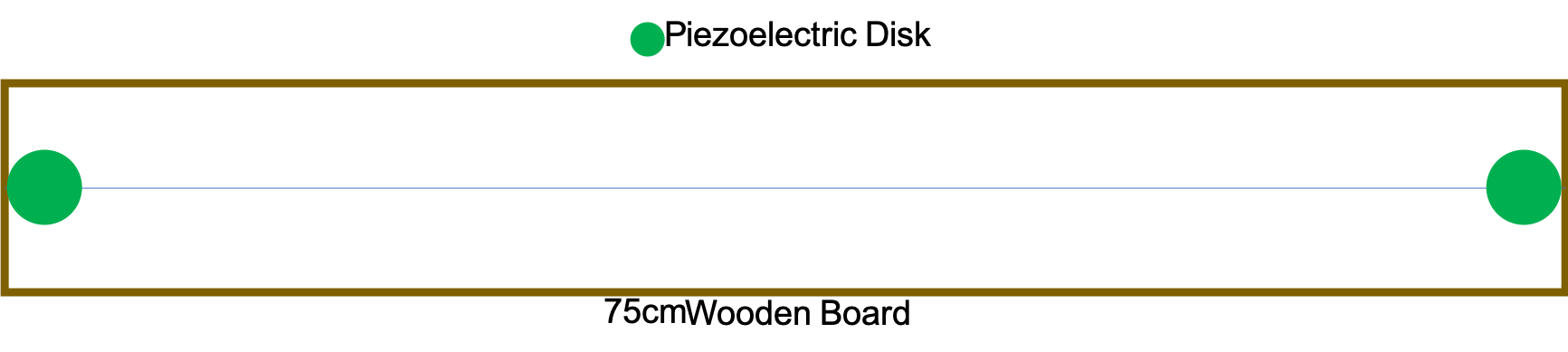} \\
  (a) Logical diagram of 1D prototype \\
  \\
  \includegraphics[width=\linewidth]{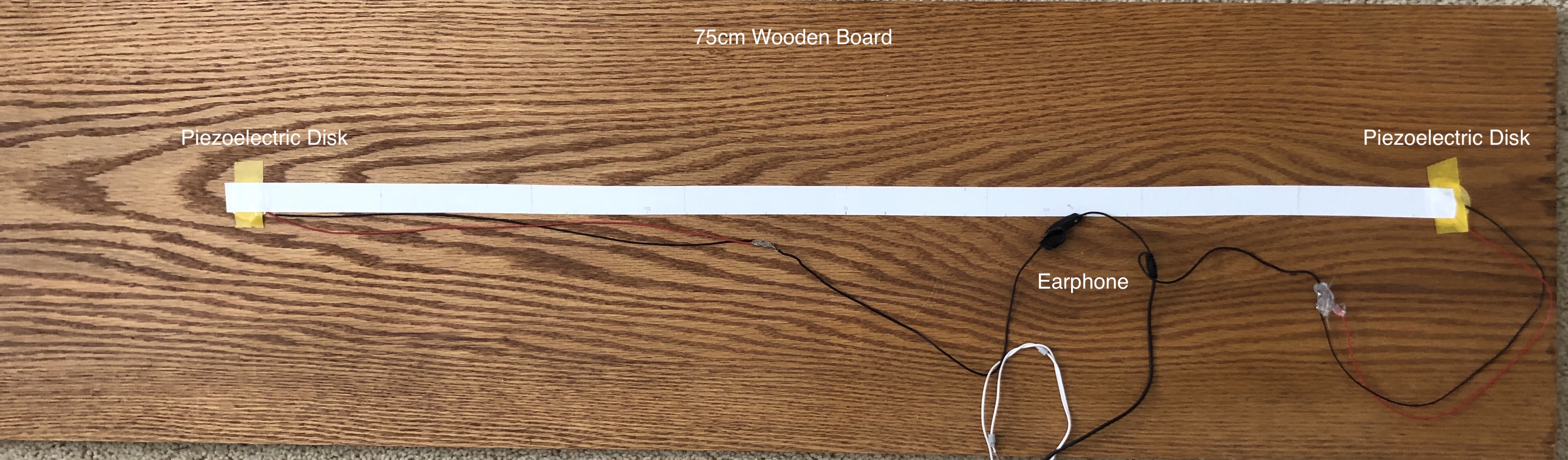} \\
  (b) Image of physical 1D prototype
  \end{tabular}
  \caption{Prototype to evaluate accuracy of \sys with a single sensor pair.}
  \label{fig:1-D-testbed}
\end{figure}

\subsection{Baseline Test for Linearity}
\label{sec:1dtest}

To evaluate the feasibility of \sys, I began with a simple experiment to measure one dimensional accuracy, only using two piezoelectric disks. The configuration of this prototype used for this experiment is shown in Figure~\ref{fig:1-D-testbed}.
I tapped every 2.5 centimeters only along a line between the two microphones. This simplified the analysis of the data since the measured time difference would simply be proportional to the progress along the line.  At each location, I tapped ten times to determine how consistently the program readings were for specific locations.

\begin{figure}[t]
  \centering
  \includegraphics[width=\linewidth]{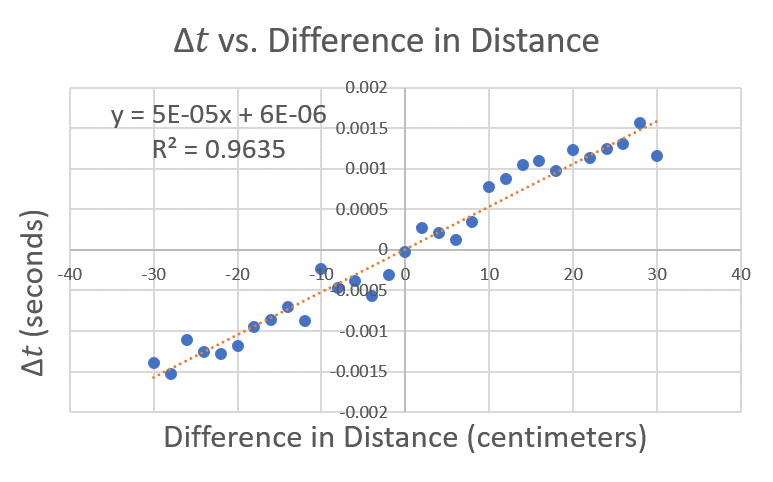}
  (a) Plot of the time difference between the two piezo disks as the tap location changes
  \includegraphics[width=\linewidth]{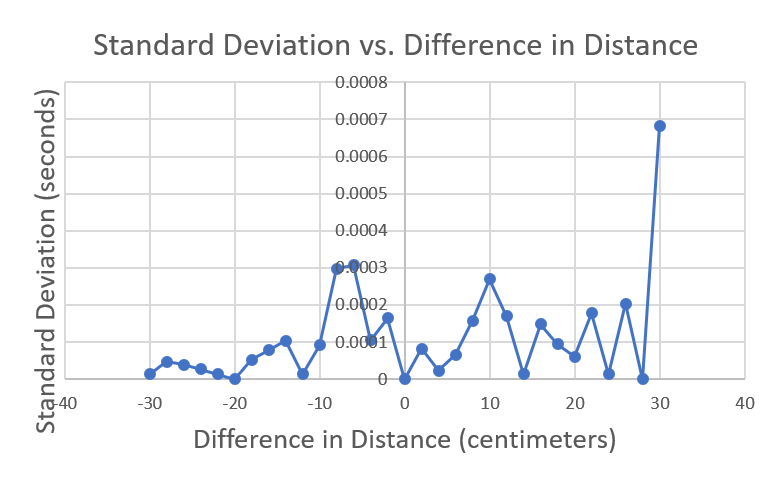}
  (b) Standard deviation at each tap location
  \caption{Test of one dimensional accuracy at 44100Hz}
  \label{fig:1-D-44100}
\end{figure}

Figure~\ref{fig:1-D-44100} (a) shows that the time difference of arrival (TDOA) follows linear relationship to the difference in distance between two microphones. This is expected because the speed of sound should be a constant through this material. The correlation coefficient, or \(R^2\) value, was high, showing that the data has a very strong linear correlation. Figure~\ref{fig:1-D-44100} (b) shows the standard deviation at each location of data collection. While the standard deviation may seem small, in relation to the expected time difference, it was very high. The variability of the readings was relatively high compared to the magnitude of the readings. This produces inaccurate distance estimates. 

\subsection{Impact of sampling frequency on accuracy}
\label{sec:sampling}

The source of the inaccuracy in the above experiements is the fact the speed travels very quickly and the resulting TDOA is very small. For example, the default configuration of the audio device is to sample at 44100 Hz. Even traveling through air, sound travels approximately 0.78cm in 1/44100 seconds. In a hard surface like a wooden board, it travels even more quickly.  

\begin{figure}[t]
  \centering
    \begin{tabular}{c}
  \includegraphics[width=\linewidth]{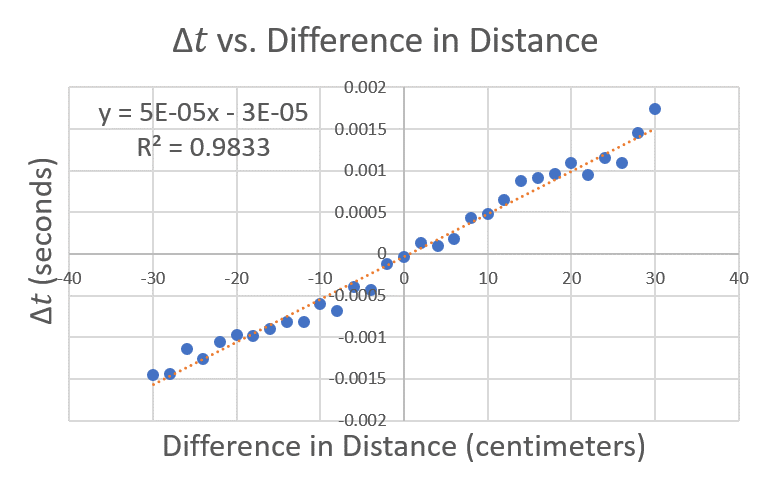} \\
  (a) Plot of the time difference between the two \\
   piezo disks as the tap location changes \\
  \\
  \includegraphics[width=\linewidth]{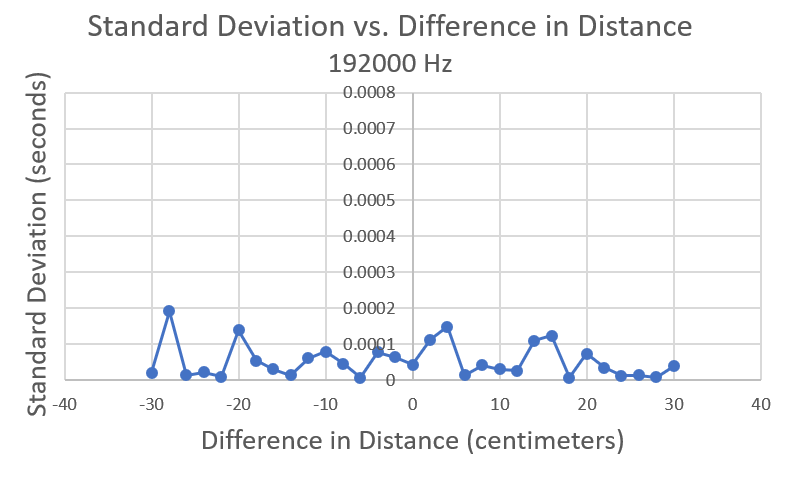}\\
  (b) Standard deviation at each tap location
  \end{tabular}

  \caption{Test of one dimensional accuracy at 192000 Hz.}
  \label{fig:1-D-192000}
\end{figure}

To address this issue, I explored the use of higher sampling frequencies. I altered the sampling rate from 44100 Hz to 192000 Hz. This significantly reduced the variability of the readings. This can be seen by the much lower standard deviation values in Figure~\ref{fig:1-D-192000} (b). The readings were more accurate and provided a better \(R^2\) value as well, as shown in Figure~\ref{fig:1-D-192000} (a).

\begin{figure}[t]
  \centering
  \begin{tabular}{c}
     \includegraphics[width=\linewidth]{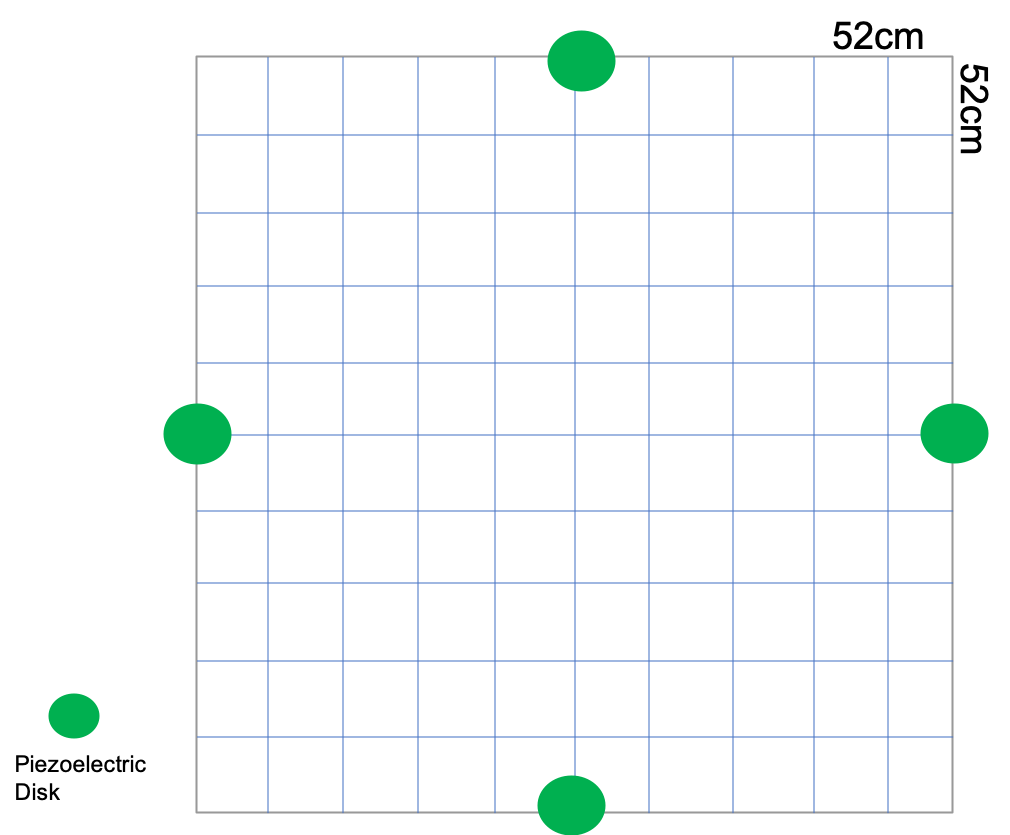} \\
(a) Logical diagram of 2D prototype \\
\\
  \includegraphics[width=\linewidth]{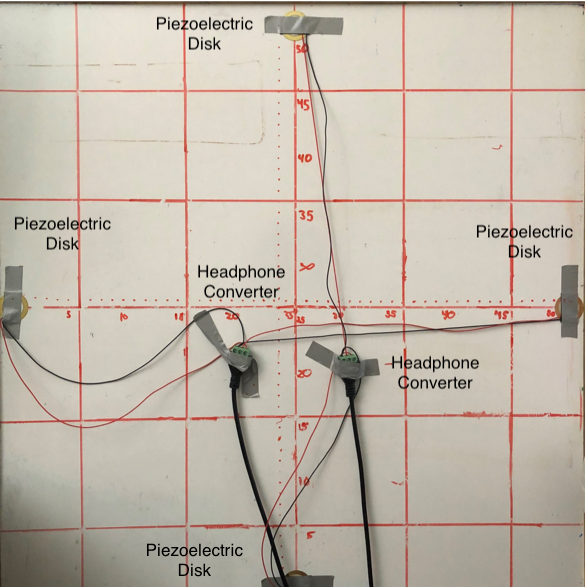} \\
(b) Image of physical 2D prototype
  \end{tabular}
  \caption{Prototype to evaluation accuracy of \sys with a two sensor pairs for $(x,y)$ tap localization.}
  \label{fig:2-D-testbed}
\end{figure}

\subsection{Surface Calibration} 
\label{sec:speedsound}

In order to make the algorithm function on multiple surfaces with varying acoustic properties, I needed to create a system for calibrating to a surface. This experiment was executed on a 2D prototype depicted in Figure~\ref{fig:2-D-testbed}. This prototype has 2 pairs of piezoelectric sensors to enable full 2D localization of taps. The goal was to have \sys produce $(x,y)$ coordinate pairs on this board that represented absolute measurements using centimeters. To achieve this goal, I needed to measure the speed of sound in the board's material. 

\begin{figure}[t]
  \centering
  \begin{tabular}{c}
  \includegraphics[width=\linewidth]{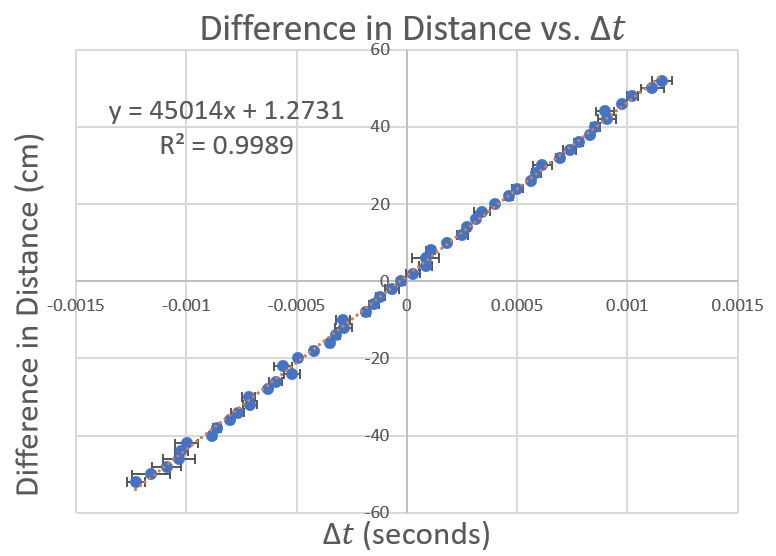}\\
  (a) Data for the x direction\\
  \\
  \includegraphics[width=\linewidth]{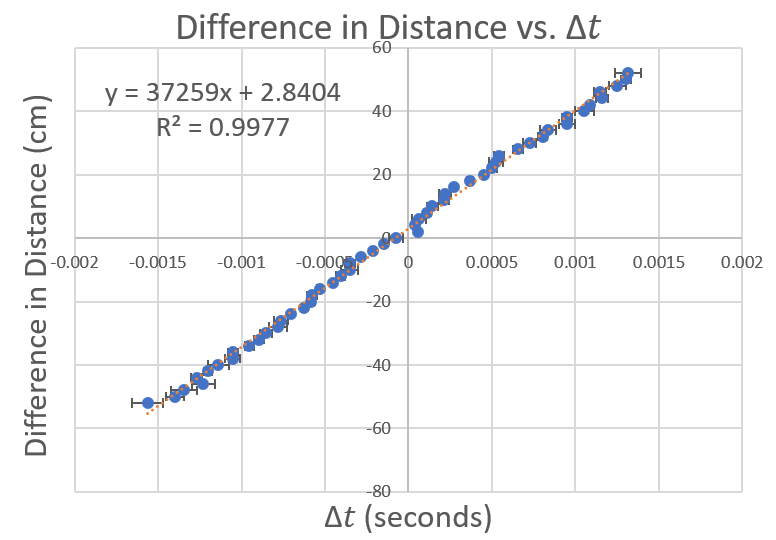}\\
  (b) Data for the y direction
\end{tabular}

  \caption{Plot of the time difference between the two piezo disks as the tap location changes. Error bars on the graph represent standard deviation.}
  \label{fig:2-D-calibration}
\end{figure}

For the calibration, I recorded 10 taps at each centimeter along the x and y axis of my board and recorded the difference in time to the sensor pair along that axis. 
Figure~\ref{fig:2-D-calibration} shows a plot of the difference in distance on the y-axis and the difference in time on the x-axis. I performed a linear fit to the data on this graph. Note that the slope of the linear trendline is the speed of the sound through the material in centimeters per second. Surprisingly, the speed of sound was different in the x and y direction - 45014cm/s in the x direction and 37259cm/s in the y direction. I conjecture that this is due to the inconsistencies in the material. For example, it is known that the grain of wood can impact the propagation of sound~\cite{acoustics}.

\subsection {Overall accuracy in two dimensions}
\label{sec:accuracy}

\begin{figure}[t]
  \centering
  \includegraphics[width=\linewidth]{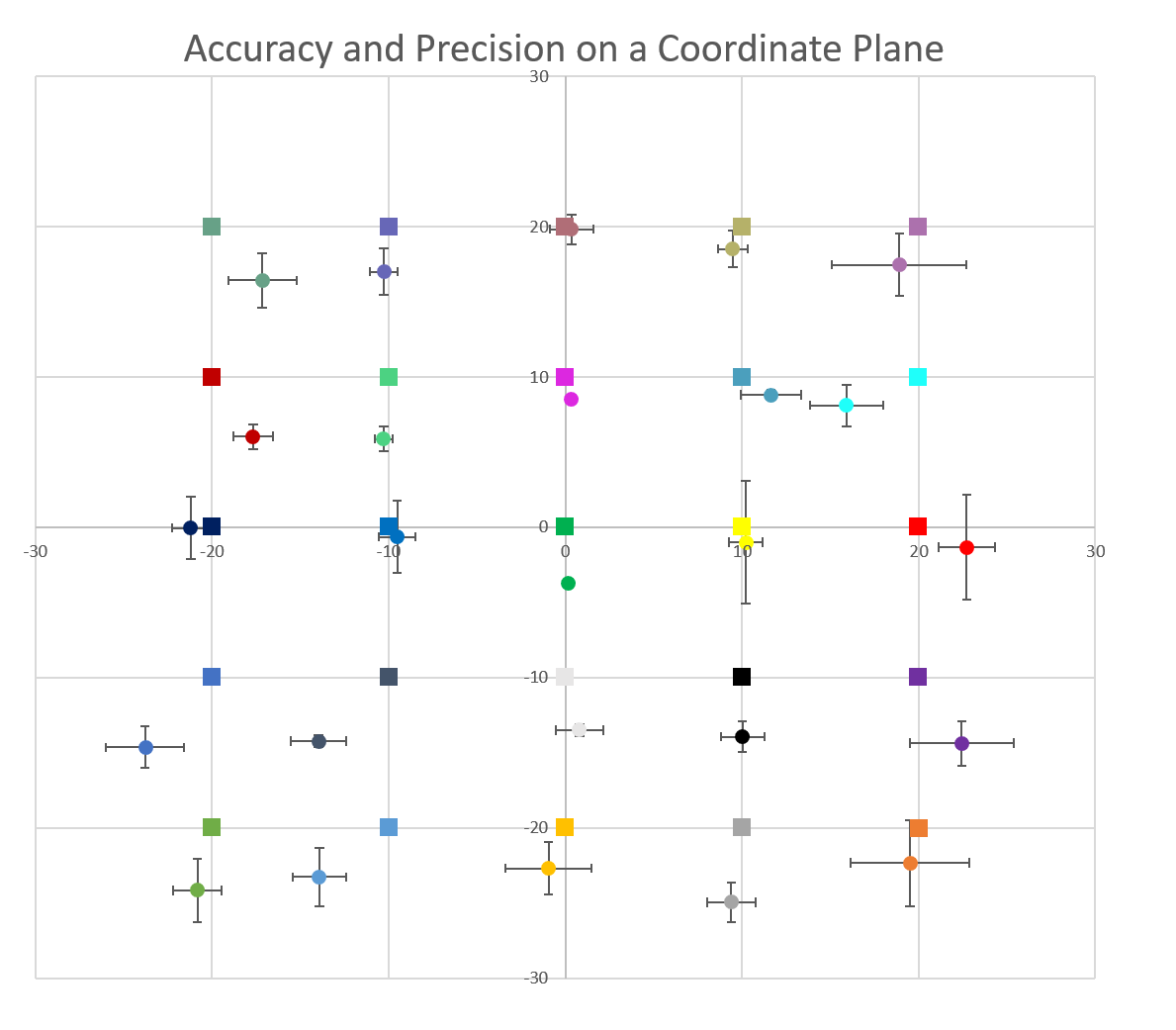}
  \caption{Two dimensional coordinate plane with tap locations every 10 cm. Error bars represent standard deviation. The squares represent the true location of the taps. Data points are colored to match their corresponding square's color.}
  \label{fig:2-D-xy}
\end{figure}

The final test to analyze the accuracy of \sys was to do a test along a two dimensions coordinate plane. This allowed me to analyze the accuracy and precision of the data when the x-coordinate and y-coordinate system is combined. I tapped 10 times every 10cm along the coordinate plane to test the accuracy and precision after combining x and y data. Accuracy is the deviation of the data from the colored square in Figure~\ref{fig:2-D-xy}. Precision is represented by the error bars that indicate standard deviation. The data shows that tap detection was highly accurate and precise because of closeness of the data to the expected position and the small error bars. Overall, the average of the absolute value of the error is 1.45cm in the x direction and 2.72cm in the y direction. 
In Figure~\ref{fig:2-D-xy}, it is noticeable that the lower half of the board has y-coordinate estimates that are biased lower than the actual tap location. I conjecture that this is a result of non-uniform sound propagation in the underlying material. This could be caused by factors such non-uniform density of the material, or defects such as cracks or knots in wood~\cite{acoustics}. 

As mentioned in Section~\ref{sec:require}, my goal was to get accuracy in the range of that needed by motor impaired users - i.e. around 2cm. The system prototype is close to achieving this target and there are promising ways to improve accuracy further as discussed in Section~\ref{sec:discussion}.

\section{Discussion}
\label{sec:discussion}

The results shared in the previous section indicate that acoustic multilateration is a viable method for touch localization. Furthermore, the project proves that a portable, accurate, and inexpensive solution can be created to address the issue of small touch interactive surface area. No prior instrumentation is necessary, allowing for an easy to implement solution. The issue of varying surfaces is addressed though a calibration system that calculates the speed of the acoustic impulse through the material. Although the project met the main constrains defined at the beginning of the paper, there are some limitations. These limitations include the following.
\begin{itemize}
    \item \textbf{Surface Material.} The reliability of the data varies slightly by surface based on its acoustic properties. While the calibration does help, there is still lower accuracy. This may be because the sound does not travel in a perfectly linear fashion in some materials. The issue may be caused by factors such as metal supports under a table or warps in wood. This could be fixed by computing a non-linear model of the sound wave through the surfaces or by using machine learning techniques to identify a piece-wise linear fit for sound propagation speeds across the material surface.
    \item \textbf{Detecting the start of the acoustic impulse.} Currently, the software uses an amplitude over a certain value to detect a tap. However, based on the loudness of the tap, this can vary. This causes error because when one microphone is further away from another, the sound is less loud to that microphone (lower amplitude values). One way to fix this is to match up the shape of the amplitude to find the actual start. Using this approach, \sys may also be able to support interactions beyond just taps. For example, \sys may be able to track drawing operations by continuously matching the sounds shapes heard across different microphones. 
    \item \textbf{Streamlining software.} Currently, the software is split into two separate python programs with most of the analysis done later. In the future, the software could be created into one single program that does all of the computation at once.
\end{itemize}

\section{Related Work}

As mentioned earlier, multilateration is not a new concept. It has been used in air traffic control systems and pre-GPS systems for localization. 

The Toffee system from CMU \cite{toffee} uses similar acoustic TDOA methods to estimate the angle of a tap relative to the device to create virtual buttons on large surfaces. \sys differs from this project as it aims to accurately identify the exact location of a tap anywhere along a coordinate plane with high accuracy, allowing for a much more natural mapping of existing application designs to arbitrary surfaces.

Accurate acoustic-based localization has been used in a variety of other systems. For example, many smart whiteboard systems (e.g.,~\cite{whiteboard}) localize pens using an ultrasonic transmitters located in the pen along with microphones embedded in the whiteboard. Similarly, past research (e.g., Cricket from MIT \cite{cricket}) has explored the use of ultrasonic transmitters embedded in the environment to help mobile devices localize themselves. These systems differ from \sys in that they all use active transmitters to help in localization. As a result, they either require power or infrastructure deployed in advance. In addition, using natural sounds such as taps prove more difficult to localize accurately. 

\section{Conclusion}

The results of the study prove that acoustic localization can be used to detect touch interaction on large surfaces with high accuracy. This is a promising first step towards creating a simple way to make any surface touch-interactive.

The system created in this project fulfills the requirements.

\begin{itemize}
    \item It is able to transform any surface into a touch interactive surface. No prior installation of materials or change to the room is required.
    \item It is small and easy to implement as it only requires inexpensive piezoelectric disks
    \item It is able to accurately identify the origination of a tap.
\end{itemize}

Some of the next steps are to test on even larger surfaces and create specific zones where a user can tap and see how small the zones can be and still maintain accuracy.

%

\bibliographystyle{ACM-Reference-Format}
\bibliography{_references}

\newpage
\onecolumn
\appendix

\section{Code}
\label{sec:code}

\subsection{Horizontal Left-Right Sensors}

Below is the code for collecting data from the left and right sensors. Note that the data collection process for the other two sensors is similar and omitted for space. 

\begin{lstlisting}[language=Python]
try:
    import os
    import pyaudio
    import numpy as np
    import pylab
    from pylab import *
    import matplotlib
    import matplotlib.pyplot as plt
    from scipy.io import wavfile
    import time
    import sys
    import seaborn as sns
    import threading
    import logging
    import math
except:
    print ("Something didn't import")
## open('leftsamples.txt', 'w').close()
## open('rightsamples.txt', 'w').close()

hit = False
hit1 = False
hit2 = False
hit3 = False
done = False
counter=0
i=0
FORMAT = pyaudio.paInt16 # We use 16bit format per sample
CHANNELS = 2
RATE = 192000 #192000
CHUNK = 8192 # (8192) 1024bytes of data read from a buffer
RECORD_SECONDS = 0.1
WAVE_OUTPUT_FILENAME = "file.wav"
left_channel = 0
right_channel = 1

audio = pyaudio.PyAudio()

# start Recording
stream = audio.open(format=FORMAT,
                    channels=CHANNELS,
                    rate=RATE,
                    input_device_index = 1,
                    input=True)

global keep_going
keep_going = True

def hit_test (piezo, amplitude): 
    for x in range(len(piezo)):
        if abs(piezo[x]) >= amplitude:
            return x
    return False    
    
def detect_tap_lr() :
    global hit
    global hit1
    global ltime
    global rtime
    global ltime2
    global rtime2
    global location
    global counter
    global done

    if done==False:
        if hit_test(left_samples, 1000) != False :
            ltime = ((hit_test(left_samples, 500))/RATE)
            hit = True
        if hit_test(right_samples, 1000) != False :
            rtime = ((hit_test(right_samples, 500))/RATE)
            hit1 = True
        if hit==True and hit1==True:
                list = [ltime, rtime]
                if list.index(min(list)) == 0 :
                    ltime2=0
                    rtime2=rtime-ltime
                    location=-22248.5249228*rtime2+26.4164968363
                if list.index(min(list)) == 1 :
                    ltime2=ltime-rtime
                    rtime2=0
                    location=22557.1202678*ltime2+26.6288158189
                hit = False
                hit1 = False
                done=True
                counter=0
                print(ltime2, rtime2, location)
                print(hit_test(left_samples, 500), hit_test(right_samples, 500))
                print("good")

##                with open('leftsamples.txt','ab') as f:
##                    np.savetxt(f, left_samples, fmt='%5d', delimiter=',')
##                with open('rightsamples.txt','ab') as f:
##                    np.savetxt(f, right_samples, fmt='%5d', delimiter=',')
                    
    else:
        if counter>4:
            done=False
        counter = counter+1

# Open the connection and start streaming the data
stream.start_stream()
print ("\n+---------------------------------+")
print ("| Press Ctrl+C to Break Recording |")
print ("+---------------------------------+\n")

# Loop so program doesn't end while the stream callback's
# itself for new data
while keep_going:
    try:
        # When reading from our 16-bit stereo stream, we receive 4 characters (0-255) per
        # sample. To get them in a more convenient form, numpy provides
        # fromstring() which will for each 16 bits convert it into a nicer form and
        # turn the string into an array.
        raw_data = stream.read(CHUNK) # always read a whole buffer.

        samples  = np.fromstring(raw_data, dtype=np.int16)
        # Normalize by int16 max (32767) for convenience, also converts everything to floats
        # normed_samples = samples / float(np.iinfo(np.int16).max)
        # split out the left and right channels to return separately.
        # audio data is stored [left-val1, right-val1, left-val2, right-val2, ...]
        # so just need to partition it out.
        left_samples = samples[left_channel::2]
        right_samples = samples[right_channel::2]

        detect_tap_lr()

    except KeyboardInterrupt:
        keep_going=False

# Close up shop (currently not used because KeyboardInterrupt
# is the only way to close)
stream.stop_stream()
stream.close()

audio.terminate()


\end{lstlisting}

\end{document}